\documentclass[aps,letter]{revtex4}

\usepackage{graphicx}
\usepackage{amsmath,latexsym,amssymb}

\begin{document}

\title{Compression modulus of macroscopic fiber bundles.}

\author{P. Beckrich$^{1}$, G. Weick$^{2}$, C. M.
Marques$^{1}$, T. Charitat\thanks{to whom correspondence should be addressed at charitat@ics.u-strasbg.fr}$^{3}$}

\affiliation{$^{1}$ LDFC, UMR 7506, CNRS-ULP, 3 rue de l'Universit{\'e}, 67084
Strasbourg Cedex, France,}
\affiliation{$^{2}$ IPCMS, UMR 7504, CNRS-ULP, 23 rue du
Loess, 67034 Strasbourg Cedex 2, France,} 
\affiliation{$^{3}$ ICS, UPR 22, CNRS-ULP, 6 rue
Boussingault, 67083 Strasbourg Cedex, France.}


\begin{abstract}
We study dense, disordered stacks of elastic macroscopic fibers. These
stacks often exhibit non-linear elasticity, due to the coupling between the
applied stress and the internal distribution of fiber contacts. We propose a
theoretical model for the compression modulus of such systems, and
illustrate our method by
studying the conical shapes frequently observed at the
extremities of ropes and other fiber structures.
\end{abstract}

\maketitle

Threadlike objects are the constituents of many natural and synthetic
materials~\cite{safin,fiberbook}. At the microscopic level, filaments
such as actin,
microtubules and other semi-flexible polymers control not only the
elastic and viscous properties of different biological
architectures~\cite{albertsbook} but also of polymer melts, surfactant
solutions and gels~\cite{doi:bookPD}. These are collective properties
that can be
understood from the individual characteristics of the filaments like
the bending and
stretching elasticity or the friction coefficient~\cite{landau:bookTE}. At such
small scales, thermal disorder plays a key role that has been
successfully described
within the framework of statistical mechanics. For instance,  the
reptation theory
for polymer melts and solutions~\cite{doi:bookPD,degennes:bookSCPP} or the
description of rigid gel elasticity~\cite{jones:91} successfully
account for the
collective macroscopic behavior, based on a few, microscopic
parameters of the single
polymers. Surprisingly, macroscopic
threadlike systems have been much less studied, in spite of their
abundance in the
natural and synthetic realms: hair, wool, cotton and other natural
fibers or Nylon
strings, ropes and textiles or glass wool belong to this class of
systems~\cite{fiberbook}. Here also the elastic or friction
constitutive parameters
of the individual objects are well known. For instance, bending and
stretching modulii
of human hair, wool, Nylon and steel fibers can easily be found in
the literature.
But theories that predict collective fiber properties such as the
compressibility of
a hair tress, or studies that explain what forces play a role in a
disentanglement
action like combing or carding, are scarse~\cite{carding,glasswool}.
In this letter
we argue that methods developed for fiber microscopic systems~\cite{selinger}
are also useful at the macroscopic level, and draw an analogy between
the effect of
thermal fluctuations and the effect of the intrinsic disorder carried by the
spontaneous shapes of the fibers.

\begin{figure}
\includegraphics[width=75mm]{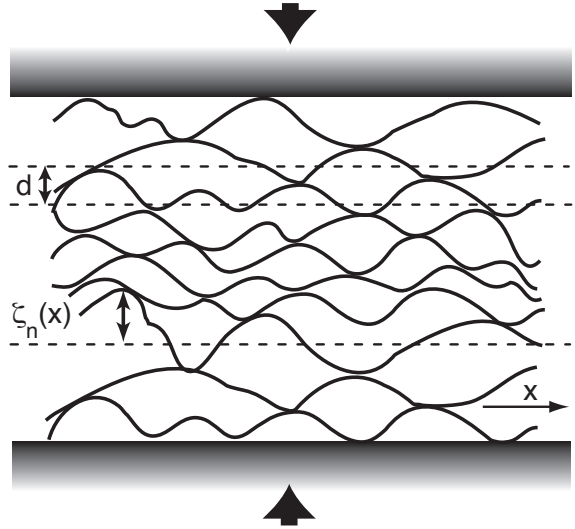}
\caption{Dense
fiber bundle. The function $\zeta_n (x)$
describes the shape of the $n$th fiber with respect to a reference
average level.}
\label{desbase}
\end{figure}

We consider a two dimensional fiber bundle as illustrated in
figure~\ref{desbase}.
Discussion in two dimensions allows for a simpler presentation of the
methods, while
retaining the main ingredients of the problem. Extension to the three
dimensional
case will be discussed later. The bundle is composed of  a large number
$\cal N$ of elastic strands of length
$L$, in a space of height $d \times {\cal N}$, where $d$ is the average
distance between
two fibers. The shape of the fiber is described by the function
$\zeta_{n}(x)$ that
measures heigth deviations from the line of average position at
height $d\times n$. This
description in the Monge parametrization is well adapted to the
shapes of interest, with a weak gradient
$d\zeta_{n}(x)/d x
\ll 1$. In
the absence of interactions between fibers, each filament
$n$ has a spontaneous shape, described by the function
$\zeta_{0,n}(x)$. It is the disorder associated with the spontaneous
shapes of the
different fibers that leads to a non-trivial compression behavior of the whole
system.  Under an external force, the spontaneous shape of the fiber
$\zeta_{0,n}(x)$ will be transformed into the actual shape
$\zeta_{n}(x)$. Within
the elastic limit, the deformation  of a segment of length
$dx$ around the natural shape will cost an elastic energy $dU_{\rm b}$,
proportional to the
bending modulus $\kappa$ and to the curvature differences,  $dU_{\rm b}=
\kappa/2 (\zeta_n''(x) -
\zeta_{0,n}''(x)
)^2 dx$.

The bending modulus $\kappa$ is a constitutive property of
the fiber, that depends on the shape of its section and on the Young
modulus of the material~\cite{landau:bookTE}. For instance, for a
cylindral filament
of radius $a$ and Young modulus $E$ one has $\kappa= E a^4/4$. The
total bending
energy of the $\cal N$ filaments can then be written as~:
\begin{equation}
U_{\rm b} =\frac{\kappa }{2} \sum_{n=1}^{\cal N} \int_0^L
dx\left[\zeta_n ''(x)- \zeta''_{0,n}(x) \right]^2.
\label{ebending}
\end{equation}
If one assumes that different fibers interact through a two body
potential $V$, then the total interaction energy $U_{\rm int}$ can be
written as
\begin{equation}
U_{\rm int} =\sum_{n,m=1}^{\cal N}
\int_0^L  dx \int_0^L  dx'
\ V\left[\zeta_m (x) + m d - (\zeta_n(x') +n d)\right].
\label{eint}
\end{equation}
In order to bring the structure of the interaction energy to a
tractable level, we follow a simple mean-field approximation, similar
to the Helfrich
treatment of thermally  activated fluid membranes and
fibers~\cite{helfrichmemb,helfrichfibers}. We assume that forces
between first-neighbors dominate the interaction energy, an exact
assumption for excluded volume potentials and a good approximation
for other short
range forces such as those given by screened electrostatic
potentials. Then the sum
of the bending~(\ref{ebending}) and interaction~(\ref{eint}) terms can be
written as the
effective energy
\begin{equation} E_{\rm eff}=\frac{\kappa }{2} \sum_{n=1}^{\cal N}
\int_0^L
dx\left[\zeta_n ''(x)- \zeta''_{0,n}(x) \right]^2 +
\frac{B}{2} \sum_{n=1}^{\cal N}  \int_0^L
dx\left[\zeta_{n+1}(x)-\zeta_{n}(x)\right]^2 \ .
\label{effnrj}
\end{equation}
Because of the weak gradients involved, inter-fiber interactions take
place only at
the same $x$ coordinate. $B$ is the strength of the mean-field
quadratic potential,
to be determined by the self-consistent prescription $B=d\times \partial^2
\left<e\right>/\partial d^2$ where $\left<e\right> = \left<E_{\rm
eff}\right>/(L
{\cal N} d)$, is the average energy density. The equilibrium state of
the fiber stack
correponds to the minimum of the effective energy
functional~(\ref{effnrj})  with
respect to the fiber configurations
$\{\zeta_n(x)\}$, given the spontaneous shapes $\{\zeta_{0,n}(x)\}$, or more
precisely, given the spontaneous curvatures $\{\zeta_{0,n}''(x)\}$.
The spontaneous
curvatures are to be treated as the random functions responsible for disorder
in the system. Averages, denoted by $\left<\right>$, will then be
performed  with
the probability associated with the spontaneous curvature distribution. Before
carrying forward the outlined procedure, it is of interest to review
results for the
thermal case, as this will help to build intuition for the key
quantities controlling the stack structure.

In thermal stacks or bundles of fibers, disorder is introduced by the
random Brownian
motion of the fibers, which would otherwise be flat, {\sl i.e.}
$\{\zeta_{0,n}(x)=0\}$. The shape of a single fiber in solution is described
by a persistent walk, a class of statistical walks where tangent correlations
vanish exponentially over a persistence length $\ell_p$, related to the fiber
rigidity through $\ell_p=\kappa/(k_B T)$. A persistent walk primarily moves
along the forward direction for a short enough distance $\ell$, but it
performs also side excursions of size $D$. From the equipartition
theorem one has
$\ell_p D^2/\ell^3 \simeq 1$. In a stack, lateral excursions are limited by
near-neighbors setting $D=d$. Typically, there is  one interaction
between fibers over a collision distance $\ell_c\simeq \ell_p^{1/3}
d^{2/3}$. The
energy density  of the thermal stack $\left<e_T\right>$,  being of the order of
$k_BT$ per collision area,
$\left<e_T\right>\simeq k_B T/(d \ell_c)$, one gets a
compression modulus:
\begin{equation}
B_T \simeq \frac{k_B T}{d^3} \left( \frac{d}{\ell_p} \right)^{1/3}=
(k_B T )^{4/3} \kappa^{-1/3} d^{-8/3},
\label{Bthermal}
\end{equation}
which holds for large enough rigidities, $\ell_p\gg d$.

In macroscopic fiber stacks, the disorder is  quenched and introduced by the
spontaneous disordered shapes  $\zeta_{0,n} (x)$. Lets consider a typical
shape of amplitude $\zeta_0$ and wavelength $q_0^{-1}$. Clearly, there are no
interactions for distances larger than the shape amplitude, $d\gg \zeta_0$. For
smaller distances the local gradients are of order of $\zeta_0 q_0$,
leading to a
collision length $\ell_c\simeq d/(\zeta_0 q_0)$. The energy density
is thus given by
$\left<e\right>\simeq \kappa \zeta_0^2 q_0^4 \ell_c/(d \ell_c)=\kappa \zeta_0^2
q_0^4/d$, independent of
$\ell_c$. It follows that the compression modulus reads:
\begin{equation}
B\simeq \kappa
    q_0^4 \left(\frac{\zeta_0}{d}\right)^2 \ .
\label{Bquenched}
\end{equation}
It is worth at this stage to stress several striking differences and
similarities
between the thermal and the macroscopic cases. First, contrary to
thermal stacks, where the compressibility is only governed by  the
persistence length
$\ell_p$ and by the distance between fibers $d$, in macrosocopic
fiber stacks the
spacial scale associated with disorder inhomogeneities plays also an
important role.
As a matter of fact, the scale for compressibility is set by the
combination $\kappa
q_0^4$:  stiffer fibers lead to harder macroscopic materials, quite
at the opposite
of thermal stacks where increasing fiber stiffness reduces fluctuations and
softens the stack. Also, while the cost of confinement for thermally
controlled fluctuations diverges as the confinement distance vanishes, only a
finite amount of energy is required to completely flatten a macroscopic
fiber.  But in both cases, the compression modulii increase with the degree
of disorder, measured by the temperature in the thermal system, and by the
square amplitude of spontaneous shapes,
$\zeta_0^2$ for the quenched bundle. And finally, the power law dependence of
the thermal modulus is larger than the macroscopic one: quenched disorder
appears to lead to more ``robust" materials as
these are less susceptible to variations in stack density. We now
return to a more
detailed calculation of the compression modulus.

\begin{figure}[h]
\includegraphics[width=80mm]{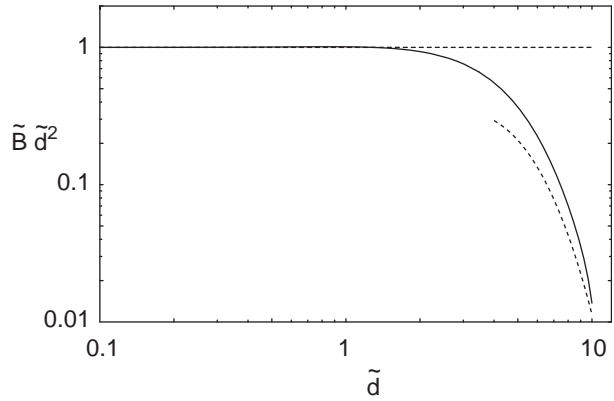}
\caption{Full line is the compression modulus $B$ of a fiber
stack {\sl vs} interfiber distance $d$. Dashed lines represent
the asymptotic behaviours at short distances,
$\widetilde B
\widetilde d^2\simeq 1$, for
$\widetilde d\ll
1$ and at large distances $\widetilde B \sim
\widetilde d \exp [ - \widetilde d\ ]$, for $\widetilde d\gg
1$. Plots are for
dimensionless quantities
$\widetilde B = B/(\kappa q_0^4)$ and $\widetilde d= d/\zeta_0$, with
$\kappa$ the fiber rigidity, $q_0$ and $\zeta_0$ the
typical wavevector
and amplitude of spontaneous shape disorder.}
\label{Bnumerique}
\end{figure}

It is convenient to perform the functional minimization of the effective
energy~(\ref{effnrj}) in the space of the eigenfunctions
$\left\{\Phi_q\right\}$
for the biharmonic operator,  $(\partial^4/\partial
x^4-q^4)\Phi_q(x)=0$. We choose
also boundary conditions ensuring that  there are no forces nor torques on the
strands extremities: $\Phi_q^{(2)}
(0)=\Phi_q^{(2)} (L)=0$ and  $\Phi_q^{(3)}
(0)=\Phi_q^{(3)} (L)=0$. The orthonormal set of
eigenfunctions can
be written as
\begin{equation}
\phi_q(x)=\frac {\cosh(\alpha_p)-\cos(\alpha_p)}
{\sinh(\alpha_p)-\sin(\alpha_p)}
\left[\sinh\left(\alpha _p \frac xL\right)+\sin\left(\alpha _p \frac
xL\right)\right]-
\left[\cosh\left(\alpha _p \frac xL\right)+\cos\left(\alpha _p \frac
xL\right)\right] \ ,
\label{basephiq}
\end{equation}
where the numerical coefficients $\alpha_p$ are determined from the relation
$\cos{(\alpha_p)} \cosh{(\alpha_p)}=1$. The solutions obey
approximately $\alpha_p
\equiv  q L\simeq \left(p+\frac 12\right)\pi$ with $p\in \Bbb{N}$. We
first develop
the shapes $\zeta_n(x)$ and $\zeta_{0,n}(x)$ on the basis of the
eigenfunctions~(\ref{basephiq}),
\begin{eqnarray}
\zeta _n(x)=\frac{1}{\sqrt{{\cal N}L}} \sum_{q=\frac{\pi }{2L}}^{+\infty }
\sum_{Q=-\frac{\pi }{d}}^{\frac{\pi}{d}}
\zeta _{qQ}\phi _q(x)e^{iQnd}
\label{zetadecompo}\\
\zeta _{0,n}(x)=\frac{1}{\sqrt{{\cal N}L}} \sum_{q=\frac{\pi }{2L}}^{+\infty }
\sum_{Q=-\frac{\pi }{d}}^{\frac{\pi}{d}}
\zeta _{0,qQ}\phi _q(x)e^{iQnd} \ ,
\label{zeta0decompo}
\end{eqnarray}
where $Q$ is a wavevector along the stack direction, then
minimise with respect to the coefficients $\zeta_{qQ}$ and finally  compute the
average energy density:
\begin{equation}
\left<e\right>=\frac {1}{2L{\cal N}d} \sum_{q,Q} \frac{q^4\left<{\zeta
_{0,qQ}}^2\right>B(Q)}{q^4+\frac{B(Q)}{\kappa }} \ ,
\label{nrjdensity}
\end{equation}
where $B(Q) = 2 B (1- \cos Q d)$. Equation~(\ref{nrjdensity}) shows
how the spatial
inhomogeneities of spontaneous shapes determine the energy density of
the stack: the
harmonic, elastic nature of the energy penalty for shape distortion and
interfiber interactions translates into the sole dependence of the
energy density on
the values of the second moment of the shape disorder distribution,
weighted by the
classical correlation kernel for bending elasticity. A practical consequence of
equation~(\ref{nrjdensity}) is that it allows to connect fiber geometric
information, available from a simple statistical analysis on fiber
shapes, with collective properties such as the compression modulus $B$. As an
example, we now calculate the compression modulus for monomodal structures of
uncorrelated fibers, for which the standard square deviation is written as
$\left<{\zeta _{0,qQ}}^2\right>= {\cal N} L \zeta_0^2 \delta_{qq_0}$.
In the limit where ${\cal N}\gg 1$, the dimensionless compression modulus
$\widetilde B=B/( \kappa q_0^4)$ obeys the self-consistent
differential equation:
\begin{equation} \widetilde B =
    \widetilde d  \frac{\partial
^2}{\partial {\widetilde d}^2}\frac{\ 1 \ }{2 \widetilde d}\left[ 1-\left(1+4
\widetilde B \right)^{-1/2}\right] \ ,
\label{selfforB}
\end{equation}
where $\zeta_{0}$ now sets the natural distance units,
$\widetilde d =
d/\zeta_{0}$.  At short distances, the compression modulus
is high, and
equation~(\ref{selfforB}) leads to the asymptotic form discussed in
(\ref{Bquenched}), $B = \kappa q_0^4
\zeta_0^2/d^2$. For distances larger than the natural amplitude for
spontaneous shape disorder, $\zeta_0$, the fibers don't interact
significantly, and the compression modulus vanishes exponentially $B
= \kappa q_0^4
(d/\zeta_0)\exp[-d/\zeta_0]$. A numerical solution of equation~(\ref{selfforB})
interpolating between these two regimes is presented in
figure~(\ref{Bnumerique}).

\begin{figure}[h]
\includegraphics[width=130mm]{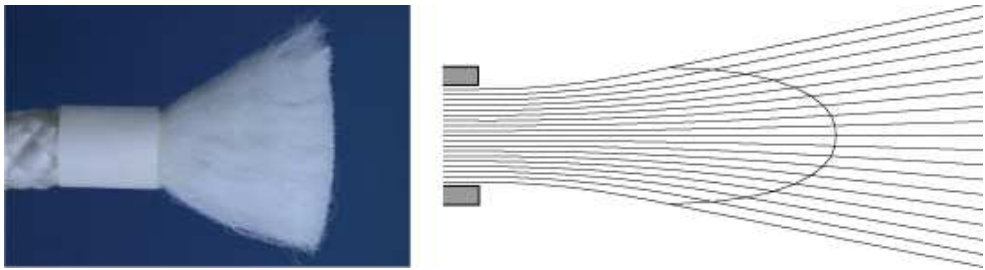}
\caption{Cones are formed at
the unconstrained ends
of ropes and other bundles of fibers, as shown on the left for a
Nylon rope. On the
right, the calculated shape in two dimensions. Here, only the average
fiber path is
represented for simplicity. The fibers only interact mutually in the expansion
zone, to the left of the curved line, and follow a straight path
otherwise. The size
of the expansion zone and  the angle of the cone are independent of
fiber rigidity,
they are fully determined  by the geometric properties of fiber disorder.}
\label{figcone}
\end{figure}

We now illustrate our results by studying the shape of the cones that
appear at the end of ropes and other fiber bundles. The cones are
generated by the
expansion of the fibers, under the repulsive field that we discussed
previously.
The repulsive field  experienced by one fiber, $U(d) = B(d) d^2/2$,
can be roughly
approximated by a step function of amplitude $\kappa q_0^4
{\zeta_0}^2/2$ and range
$\zeta_0$. The average path of one fiber is a straight,  horizontal line,
as long as it remains inside the constrained zone, as depicted in
figure~(\ref{figcone}). When it escapes confinement, the fiber
crosses an expanding
zone where it  still interacts with its neighbors. Under the
repulsive forces, the
fiber bends until it escapes the repulsive potential, for interfiber distances
larger than the range of the potential  $\zeta_0$. Bending of the
fiber is caused by two localized forces of amplitude $f = \kappa
q_0^4 \zeta_0^2/2$
and opposite signs, applied by the two closest neighbors.  After escaping the
expanding interaction zone, the fiber follows a straight line, at an angle that
depends on its distance from the center of the bundle. If one
enumerates the fibers
in the stack from $n=1$ in the center to $n={\cal N}/2$ for the
external fiber, the
following set of equations recursively determine the position
$[\ell_n,y_n]$ where a fiber escapes its neighbors potential
\begin{eqnarray}
(\ell_{n-1}-\ell_n) (3 \ell_{n-1}^2 -(\ell_{n-1} - \ell_n)^2) & =  &
\frac{6 \kappa}{f}
(y_{n-1} - (n-1)d) \\
(\ell_{n-1}-\ell_n) \ell_{n}^2 & = & \frac{2 \kappa}{f}  (y_n - n d
-(\zeta_0 -d))\ ,
\label{conerecursive}
\end{eqnarray}
where $\ell_n$ is measured from the border of the confinement zone
and $y_n$ from
the center of the stack. The distance $d$ is the interfiber separation in the
confinement zone. Equations~(\ref{conerecursive}) and the boundary
conditions $y_1
=\zeta_0$ and $\ell_{{\cal N}/2}^3= (y_{{\cal N}/2} -  d {{\cal N}/2}) 3
\kappa/f$ determine the shape to the boundary $y_n(\ell_n)$, shown also in
figure~(\ref{figcone}). The extension of the expansion zone,
$\ell_1$, depends on
the number of fibers and on the compression rate
$\alpha=(\zeta_0-d)/\zeta_0$, but not on the fiber rigidity:
$\ell_1\sim \zeta_0 {\cal N}^{2/3} \alpha^{1/3} (q_0
\zeta_0)^{-4/3}$. Similarly, the geometric shape of the cone, given
by the slope of the last fiber $\theta_{{\cal N}/2}$, does not depend on the
fiber rigidity, it is only fixed by the geometric quantites associated with
the fiber disorder, $\theta_{{\cal N}/2}\sim \alpha^{2/3} (q_0
\zeta_0)^{-4/3}$. 

\begin{figure}[h]
\includegraphics[width=55mm]{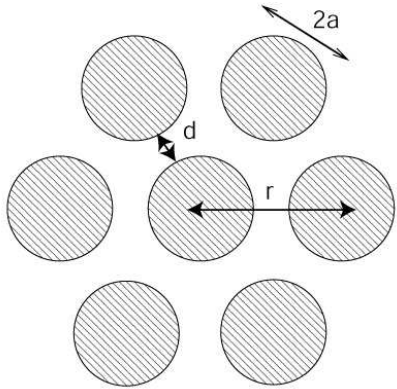}
\caption{Cross section of a three dimensional bundle in the high
density limit where
each fiber is confined in a cage defined by its first nearest neighbors.}
\label{fig3d}
\end{figure}

For high enough fiber density, the compression modulus of a three dimensional bundle can be computed in manner
similar to the two dimensional case.
Indeed, if the
density is so high as to provide a cage environment for each of the fibers, as
depicted in  figure~(\ref{fig3d}), the forces between fibers are controlled
  by excluded volume interactions between near-neighbors. In this case, we
define a two dimensional fiber density in the bundle $\sigma = {\cal N} \pi
a^2/{\cal A}$ where
${\cal A}$ is the area of the bundle section.
For a perfectly ordered
hexagonal array one would have $\sigma = 2 \pi \sqrt{3}/3  \  a^2
r^{-2}$, where $r$
is the distance between fiber centers. The maximum close packing density is
$\sigma_{\rm max} \simeq 0.91$. The relevant fluctuating distance
$d = r- 2 a$ is now a function of the fiber density $\sigma$.
In the hexagonal case, the cage would hold up to to distances $d = 2
a$ or densities
as small as $\sigma_c= \pi \sqrt{3}/24 \simeq 0.23$. In this regime,
the compression
modulus is expected to vary as $B_{\rm 3D} \sim \kappa q_0^4
\zeta_0^2 r^{-3}$. As explained above, for the non-thermal case, only a
finite energy is required to completely flatten a fiber. Thus, the
confinement  energy density is here bound to a maximum value $<e_{\rm 3D}>
\sim \kappa q_0^4
\zeta_0^2 a^{-2}$, contrary to the three dimensional thermal
situation~\cite{selinger}, where the energy density diverges as $\sim
d^{8/3}$. For densities smaller than
$\sigma_c$, the
situation is unclear, even for the thermal case. Large
lateral  excursions of the
fiber are now possible~\cite{witten}, and there is not yet, to our knowledge, a
clear criterium to determine the maximum allowed size of the
excursions. We hope to
address this question in future work.

As a summary, we have shown that stacks of macroscopic disordered
fibers behave as a
compressible material, the compression modulus being {\sl
proportional} to fiber
rigidity, at the opposite of microscopic fiber stacks where it
{\sl decreases} with fiber rigidity.  We have also shown how the compression
modulus depends on stack density and on the geometric quantities
characterizing spontaneous shape disorder. Our results allow to
predict collective properties from the individual fiber
characteristics and provide
a  new tool to interpret experiments in fiber systems. As an example of this we
studied terminal bundle cones, showing that this class of shapes
follows an universal behavior that depends only on fibers geometric features.
Extension of our results to three dimensions is straightforward in the limit of
large fiber density. The weak density limit poses an interesting
challenge for fiber
and polymer systems, both in the thermal and quenched cases.

\acknowledgments  C.M. acknowledges support from the CNRS Chemistry Department
under AIP ``Soutien aux Jeunes Equipes". We are grateful for
inspiring discussions
with F. Leroy, P. Barbarat and T. Witten.

\end{document}